\pdfoutput=1 

\documentclass{article}

\usepackage{arxiv}

\usepackage[utf8]{inputenc} 
\usepackage[T1]{fontenc}    
\usepackage{hyperref}       
\usepackage{url}            
\usepackage{booktabs}       
\usepackage{amsfonts}       
\usepackage{nicefrac}       
\usepackage{microtype}      
\usepackage{lipsum}		
\usepackage{graphicx}
\usepackage{doi}
\usepackage[table]{xcolor}

\usepackage{hyperref}       
\usepackage{url}            
\usepackage{booktabs}       
\usepackage{amsfonts}       
\usepackage{nicefrac}       
\usepackage{microtype}      
\usepackage{lipsum}
\usepackage{fancyhdr}       
\usepackage{graphicx}       
\usepackage{amsmath}
\usepackage{physics}
\usepackage{tikz}
\usetikzlibrary{quantikz}

\usepackage[table]{xcolor}
\definecolor{Gray}{gray}{0.925}

\usepackage{caption}
\usepackage{subcaption}

\usepackage{csquotes}
\usepackage{hyperref}

\usepackage{soul}
\usepackage{siunitx}
\usepackage{extarrows}

\usepackage{algorithm}

\usepackage[normalem]{ulem}               
\newcommand\redout{\bgroup\markoverwith
{\textcolor{red}{\rule[0.5ex]{2pt}{0.8pt}}}\ULon}

\graphicspath{{media/}}     

\definecolor{slcolor}{RGB}{128,64,255}

\usepackage{balance} 
\usepackage{soul}
\usepackage{tikz}
\usepackage{multirow}
\usepackage{amsthm}

\usepackage{caption}
\usepackage{subcaption}

\usepackage{algorithm}
\usepackage{algpseudocode}
\usepackage{physics}

\title{Enabling Non-Linear Quantum Operations through Variational Quantum Splines}

\author{Matteo Antonio Inajetovic\\ 
Dept. of Computer Science and Engineering\\ 
University of Bologna\\
Bologna, Italy\\
\texttt{matteo.inajetovic@studio.unibo.it}
\And
Filippo Orazi\thanks{Corresponding author}\\ 
Dept. of Computer Science and Engineering\\ 
University of Bologna\\
Bologna, Italy\\
\texttt{filippo.orazi2@unibo.it}
\And
Antonio Macaluso\\
German Research Center for Artificial Intelligence \\ DFKI\\
Saarbruecken, Germany\\
\texttt{antonio.macaluso@dfki.de}
\And
Stefano Lodi\\
Dept. of Computer Science and Engineering\\ 
University of Bologna,\\
Bologna, Italy\\
\texttt{stefano.lodi@unibo.it}
\And
Claudio Sartori\\
Dept. of Computer Science and Engineering\\ 
University of Bologna,\\
Bologna, Italy\\
\texttt{claudio.sartori@unibo.it}
}

\date{March 2023}
\begin{document}
\maketitle
\begin{abstract}
The postulates of quantum mechanics impose only unitary transformations on quantum states, which is a severe limitation for quantum machine learning algorithms. 
Quantum Splines (QSplines) have recently been proposed to approximate quantum activation functions to introduce non-linearity in quantum algorithms.
However, QSplines make use of the HHL as a subroutine and require a fault-tolerant quantum computer to be correctly implemented.    \\
This work proposes the Generalised Hybrid Quantum Splines (GHQSplines), a novel method for approximating non-linear quantum activation functions using hybrid quantum-classical computation. The GHQSplines overcome the highly demanding requirements of the original QSplines in terms of quantum hardware and can be implemented using near-term quantum computers. Furthermore, the proposed method relies on a flexible problem representation for non-linear approximation and it  is suitable to be embedded in existing quantum neural network architectures.
In addition, we provide a practical implementation of the GHQSplines using Pennylane and show that our model outperforms the original QSplines in terms of quality of fitting.
\end{abstract}


\section{Introduction}
 The ability to approximate non-linear activation functions in a quantum computer is essential to unlocking the full potential of quantum machine learning algorithms. 
 Recently, the Quantum Splines (QSplines) \cite{qsplines} adopt the \textit{B-splines} parametrization \cite{deboor} to define a linear system of equations with a block design matrix where the \textit{sparsity} is constant and depends on the degree of the polynomial fitted in each local interval. Specifically, given a sequence of knots $\xi_1, \xi_2, \cdots, \xi_T$, a line is fitted in each interval
$\left[\xi_k, \xi_{k+1} \right]_{k = 1, \cdots, T-1}$
without derivability constraints. 

\begin{equation} \label{eq:B-spline_lin_system}
\boldsymbol{\Tilde{y}} = \boldsymbol{S\beta} \rightarrow 
\begin{pmatrix}
\Tilde{y}_1 \\
\Tilde{y}_2 \\
\cdots \\ 
\Tilde{y}_{K}\\
\end{pmatrix}
=
\begin{pmatrix}
S_1 & 0 & \cdots & 0\\
0 & S_2 & \cdots & 0\\
\cdots & \cdots    & \cdots & \cdots\\ 
0 & 0 & \cdots & S_{K}\\
\end{pmatrix} 
\begin{pmatrix}
\beta_1 \\
\beta_2 \\
\cdots \\ 
\beta_{K}\\
\end{pmatrix} 
,
\end{equation}

where $\Tilde{y_k}$ represents the function evaluations in the $k$-th interval $[\xi_k, \xi_{k+1}]$, $\beta_k$s are the spline coefficients and $\boldsymbol{S}_{(2K) \times (2K)}$ is a block diagonal matrix with each block $S_k$ that representing the correspondent basis expansions of the input.  
The algorithm of QSplines considers the B-spline formulation in Eq. \eqref{eq:B-spline_lin_system} and implements it using quantum computation to approximate non-linear functions. In particular, the computation is performed in three steps. First, the HHL \cite{HHL} computes the spline coefficients $\ket{\beta_k}$ for the $k$-th interval. Second, $\ket{\beta_k}$ interacts with the quantum state $\ket{x_{k}}$ encoding the input in the $k$-th interval via quantum interference through the swap-test \cite{PhysRevLett.87.167902}. Third, $\ket{f_{k}}$ is measured and post-processed to obtain $y_{k}$. 
\\
Although QSplines allow overcoming the limitation of unitary operation on quantum states, their applicability is very limited since using the HHL as a subroutine requires a huge number of error-corrected qubits to be executed. Furthermore, in the current formulation, the final quantum state is obtained using the swap test \cite{swap}, which implies the value of the non-linear activation function is not directly encoded into the amplitude of a quantum state and needs a post-processing step to be calculated. All these factors limit the adoption of QSplines in current models of quantum neural networks, which run on a limited set of noisy qubits.

We propose the Generalised Hybrid Quantum Splines (\textit{GHQSplines}), a novel variational quantum approach for non-linear quantum activation functions based on QSplines. GHQSplines adopt the problem formulation of the QSplines \cite{qsplines} using the \textit{VQLS} \cite{vqls} as a quantum linear solver and the quantum dot product \cite{markov2022generalized} instead of the swap-test.\\ The advantages of such an approach are the following: firstly, using the VQLS, the model can be executed using near-term quantum technology. Secondly, the quantum dot product is computationally more efficient than the swap test and doesn't require a post-processing step. Furthermore, the new formulation allows obtaining an end-to-end quantum routine for any non-linear activation functions that can be adopted as a sub-routine in existing quantum neural networks.

\section{Generalized Hybrid Quantum Splines}

The original formulation of the QSplines and VQSplines relies on a block diagonal B-Spline matrix which allows decomposing the entire problem into $K$ sub-problems that can be solved independently. However, this approach does not allow the estimation of a non-linear function in a full quantum manner, which is a requirement for embedding a quantum activation function into a quantum neural network. 
For this reason, we propose the \textit{Generalized Hybrid Quantum Splines} (GHQSplines), a novel approach that relies on a more general formulation of the B-splines and allows estimating the value of a non-linear function with a single quantum circuit without problem decomposition.

Given the recursive definition of B-Spline \cite{deboor} and the related basis expansion with knots list $\xi=[\xi_1,..,\xi_i,\xi_{i+1},..,\xi_T]$,
a non linear function $f$ can be estimated using the observed values $Y=\{y_1,..,y_K\}$ given the  inputs $X = \{x_1,..,x_K\}$.
In this case, the linear system of equations describing the relation between the estimates of the activation function $Y$, the matrix $S$, and the spline coefficients $\beta$ are the following:
    \begin{equation}\label{eq:general b-spline}
    \textbf{$Y_{K \times 1}=S_{K\times K}\beta_{K\times 1}$}
        \implies
    \begin{bmatrix}
    y_{1} \\ y_{2} \\ \vdots \\ y_{K}
    \end{bmatrix}
    =
    \begin{bmatrix}
    B_{1,d}(x_1)  &\dots &  B_{l,d}(x_1)\\
    B_{1,d}(x_2)  &\dots &  B_{l,d}(x_2)\\
    \vdots  & \ddots & \vdots\\
    B_{1,d}(x_K)  &\dots &  B_{l,d}(x_K)\\
    
    \end{bmatrix}
    \begin{bmatrix}
    \beta_1 \\ \beta_2 \\ \vdots \\\beta_K 
    \end{bmatrix},
    \end{equation}
    where $d$ is the degree of the B-spline, $T$ is number of knots and $l=T-d-1$.
    
    Nonetheless, to adopt the VQLS (or the HHL) for solving the linear system of equations, the basis expansion matrix $S$ has to be hermitian and, therefore, square and non-singular. 
    The GHQSplines formulation imposes the matrix $S$ to be Hermitian, i.e.,  $K = l = T -2$, and adopts a quantum linear solver to find the set of optimal parameters $\ket{\beta}$\footnote{It is also possible to define the Hermitian matrix $H$ from $S$ as $H=\begin{pmatrix}
0 & S \\
S^\dag & 0 
\end{pmatrix}$.}. Assuming to fit a linear function in each interval (i.e., $d=1$), the linear system of the GHQSplines is: 
    
    \begin{equation}
    \begin{bmatrix}
    y_{1} \\ y_{2} \\ y_{3} \\ y_{4} \\ \vdots \\ y_{k-1} \\ y_{K}
    \end{bmatrix}
    =
    \begin{bmatrix}
    1 & 0  & \dots & \dots & 0 & 0\\
    0 & 1-x_2  & x_2 & \dots & \dots & 0\\
    \dots & 0 & 1- x_3 & x_3 & \dots& \dots  \\
    \dots  & \dots & 0 & 1-x_4 & \dots & \dots \\
    \dots  & \dots  & \dots  & \dots & \dots & \dots\\
    0 & \dots & \dots & \dots & 1-x_{K-1} & x_{K-1} \\
    0 & 0 & \dots &  \dots& 0 & 1 
    \end{bmatrix}
    \begin{bmatrix}
    \beta_1 \\ \beta_2 \\  \beta_3 \\ \beta_4 \\ \vdots \\ \beta_{K-1} \\\beta_K 
    \end{bmatrix}.
    \label{v2system}
    \end{equation}
    which leads to the following quantum linear system of equations:
    \begin{align}\label{eq:GHQSplines linear system}
        S \ket{\beta} = \ket{Y}
    \end{align}    
    where $\ket{\beta}$ and $\ket{Y}$ are two quantum states that encode in the amplitudes the vectors $Y$ and $\beta$. 
    Importantly, the normalization constraint of quantum states  $\ket{\beta}$ and $\ket{Y}$ imposes
    rescaling the values of the target variable such that $\braket{Y}{Y}=\braket{\beta}{\beta} = 1$.

    With a particular focus on the VQLS as a quantum linear solver, the size of the linear system in Eq. \eqref{eq:GHQSplines linear system} is $K$ which gives us the number of qubits required to find the optimal coefficients. Precisely, since the VQLS requires the vectors to be encoded as quantum state, the number of qubits scale logarithmically in the number of knots $K$, which is directly related to the quality of the fitting of the curve (the higher is $K$, the better the fitting is). This exponential scaling with respect to the number of inputs is a significant improvement when compared to other quantum approaches for quantum activation functions \cite{qaf_qnn} whose number of qubits scales linearly with the size of the inputs. \\
\begin{figure}[t]
    \centering

    \includegraphics[width =\textwidth]{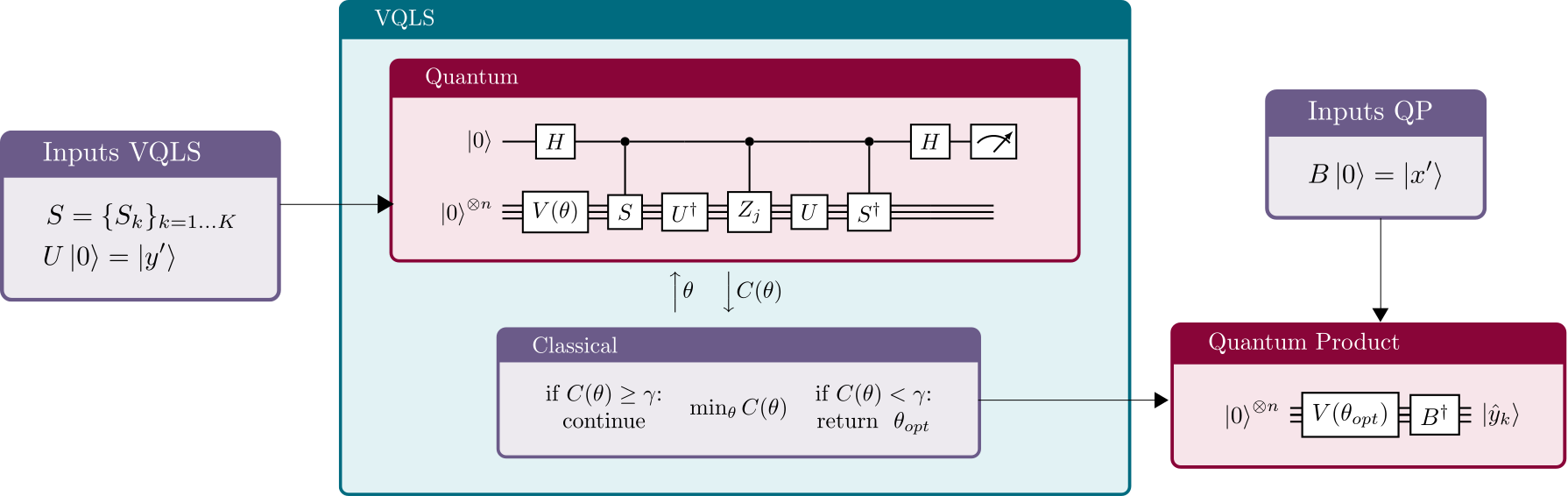}
    \caption{GHQSplines architecture. The quantum part of the VQLS is optimized classically to obtain $\theta_{opt}$ that is then used in the quantum inner product to compute $\ket{\hat{y}_k}$. All $y'_k \in y$ can be approximated with the results of a single optimization of the VQLS. }
    \label{fig:GHQSplines}
\end{figure}
    Given the linear system in Eq. \eqref{eq:general b-spline}, the GHQSplines proceeds in two steps. Firstly, the coefficients $\ket{\beta'}$ are generated using the VQLS (or the HHL): 
    \begin{equation}
        \ket{\beta'} = VQLS(S,Y) = \sum_{i=1}^K \beta_i\ket{i}.
    \label{vqlsbetav2}
    \end{equation}
    Thus, quantum state $\ket{\beta'}$ interacts via interference with $\ket{x'}$ representing the basis expansion of $x$ by means of the quantum inner product. 
\\
In the case of GHQSplines, the VQLS circuit requires $n+1$ qubits, where $K=2^n$ is the number of knots. Therefore, the Hadamard test \cite{hadamardtest} employs $n$ qubits to implement the operators ($V(\theta)$,$A$, and $U$) and one for the ancilla qubit. Furthermore, if $d=1$, we end up with a diagonal block matrix which allows decomposing the matrix $S$ (Eq. \eqref{eq:B-spline_lin_system}) in terms of quantum gates in such a way to define the quantum circuit of the VQLS efficiently. 
Specifically, we can act independently on each interval to define the $k$-th matrix decomposition as follows: 
    \begin{equation}
            S_k =
            \begin{bmatrix}
                1-a & a\\
                0 & 1-b\\
            \end{bmatrix}
            = 
            \sum_{l=0}^3 A_lc_{k,l}
            = Ic_{k,0} + Xc_{k,1} + Zc_{k,2} + R_y(3\pi)c_{k,3}  
        \end{equation}
        where the coefficients of the linear combination are computed as:
        \begin{equation*}
            c_0=1-a/2-b/2; \hspace{2em} c_1=a/2; \hspace{2em} c_2=(b-a)/2; \hspace{2em} c_3=a/2
        \end{equation*}
    This method allows the $S$ matrix to be efficiently decomposed and obtain the linear combination of quantum gates required for the VQLS (\cite{vqls}). The operator $U$ is realized by amplitude encoding state preparation \cite{mottonen}. Still, the normalization of $Y$ is required. With this new formulation, we are able to solve a singular linear system and encode all the spline coefficients $\beta$ in a unique quantum state by implementing the Ansatz only once.
    Subsequently, this state is used to compute the inner product with the $k$-th row of the matrix S (encoded through the routines $B_i$) and return the $\hat{y}_k$ estimates describing $\hat{Y}$.
    The workflow of GHQSplines is depicted in Figure   \ref{fig:GHQSplines}.

\subsection{Evaluation}

We implement the GHQSplines to approximate three non-linear activation functions (\textit{sigmoid}, \textit{elu}, \textit{relu}) and the \textit{sine} function. The functions are normalized in the interval $[0,1]$ since the input features and the value of the non-linear activation function needs to be encoded into the amplitude of a quantum state. In order to perform a fair comparison between the models, we use the 
Normalized Root Mean Squared Errors (NRMSE), defined as:
\begin{equation}
    \mathbf{NRMSE} = \frac{\sqrt{N^{-1}\sum_{i=0}^N (\hat{y}_i - y_i)^2}}{y_{max} - y_{min}}.
\end{equation}
The NMRSE allows for a robust comparison independently of the scale of the target variable and the number of knots. Thus it allows a fair comparison between the GHQSplines and the original formulation of QSplines. 
 \begin{figure}
    \centering
        \includegraphics[scale=.23]{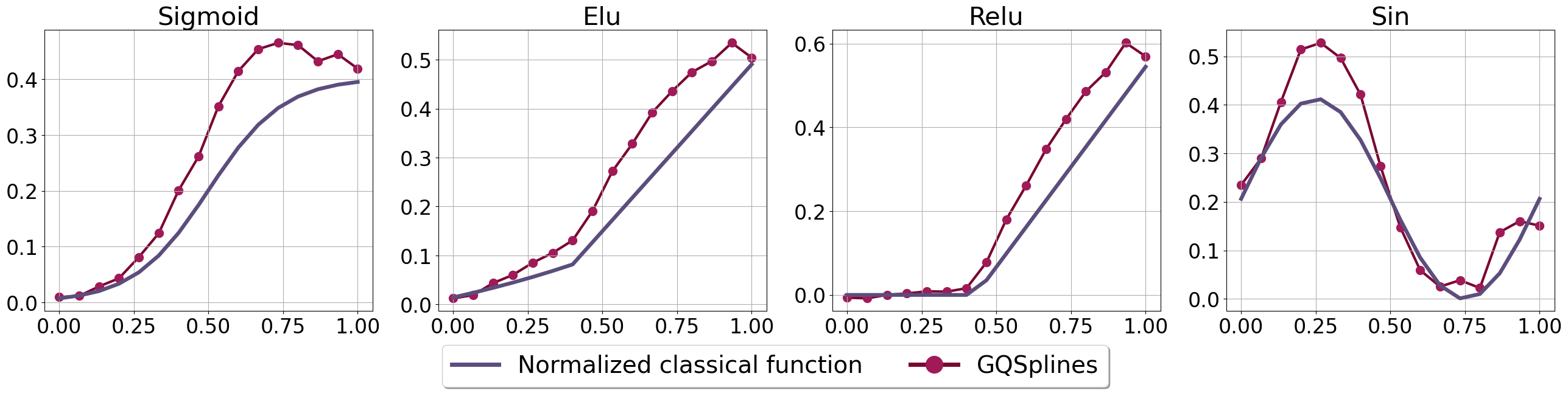}
    \caption{GHQSplines experimental results. The model is able to approximate and emulate the trend of all 4 normalized functions. Each point is computed as the product of the $j$-th row of $S$ and the B-spline coefficients $\ket{\beta'}$. }
    \label{fig:GVQS}
\end{figure}\\
 The experimental evaluation of GHQSplines is performed on each function and the results for a $4$-qubit implementation are depicted in Figure \ref{fig:GVQS}. 
We can observe that there is a slight tendency of the GHQSplines to overestimate the target function. Nevertheless, the method can capture the non-linearity of the curves, especially in the case of the activation functions. 
Table \ref{tab:my_label} reports the results of the model in terms of NRMSE and a comparison with the original QSplines function. 
We can observe that the proposed method outperforms the original QSplines on the three non-linear activation functions where both methods are tested.
\begin{table}[H]
\centering   
\begin{tabular}{|m{60pt}|m{30pt}<{\centering}|m{40pt}<{\centering}|m{40pt}<{\centering}|m{40pt}<{\centering}|m{40pt}<{\centering}|}
\hline
\hline
 \multicolumn{1}{c}{ \textbf{Model}} &\multicolumn{1}{c}{  \textbf{Knots} }         &    \multicolumn{1}{c}{ \textbf{Elu}} &\multicolumn{1}{c}{    \textbf{Relu}} &  \multicolumn{1}{c}{ \textbf{Sigmoid}} &\multicolumn{1}{c}{    \textbf{Sin}}\\
\hline
\hline
 \rowcolor{Gray} Qsplines & 20  & 0.4874   & 0.5240 & 0.1589   &  ---   \\\hline
                GHQSplines & 16  & \textbf{0.0126} &\textbf{ 0.0111}  & \textbf{0.0156} & 0.0099 \\\hline
\end{tabular}
\vspace{.5em}
    \caption{NRMSE on each function for the proposed model and the baseline. The best approximation for each function is highlighted, and we can see that our model provides a considerable increase in performance with respect to QSplines. }
    \label{tab:my_label}
\end{table}
\section{Conclusion}
Quantum Machine Learning (QML) has recently attracted ever-increasing attention and promises to impact various applications by leveraging quantum computational power and novel algorithmic models, such as Variational Algorithms.
One of the major issues in building a complete quantum neural network is the limitation of unitary, and therefore linear, operations on quantum states. In this work, we move toward a non-linear approximation of quantum activation functions using parametrized quantum circuits. In particular, we showed that it is possible to circumvent the constraint of unitarity in quantum computation by presenting an efficient version of the QSplines, whose implementation falls within the context of fault-tolerant quantum computation.
 GHQSplines are formulated and discussed as a variational approach to QSplines. The benefit of this new formulation lies in the ability to be generalizable with respect to the structure of the spline matrix. The GHQSplines adopt a new basis expansion matrix formulation, avoid the problem decomposition, and allow for tackling the problem of the matrix inversion in an end-to-end manner, with one single linear system and the number of qubits that scales logarithmically with the number of knots. Furthermore, the GHQSplines are more efficient with respect to the number of qubits required with respect to existing quantum approaches for quantum activation functions. 
Experiments showed that GHQSplines outperform QSplines and are indeed able to capture non-linearity. 
Future work will be dedicated to embedding the GHQSplines as a subroutine in existing quantum neural networks \cite{qslp, macaluso2020variational, amsdottorato9791}to leverage their full power.

\section*{Acknowledgments}
This work has been partially funded by the German Ministry for Education and Research (BMB+F) in the project QAI2-QAICO under grant 13N15586.

\bibliographystyle{unsrt}
\bibliography{reference}  

\end{document}